\documentclass[prc,
 preprint,
 showpacs, amsmath,amssymb, aps, floatfix]{revtex4}
 \usepackage{graphicx}
 \usepackage{bm}
 \usepackage{amssymb}
 \usepackage{amsmath}

\begin{document}
\title{
An in-medium full-folding model approach 
to quasielastic (p,n) charge-exchange reactions
}
\author{H. F. Arellano}
\email{arellano@dfi.uchile.cl}
\homepage{http://www.omp-online.cl}
\affiliation{Department of Physics - FCFM, University of Chile \\
             Av. Blanco Encalada 2008, Santiago, Chile}
\author{W. G. Love}
\email{wglove@physast.uga.edu}
\affiliation{Department of Physics and Astronomy,
         University of Georgia,
         Athens, GA-30602, U.S.A.}
\date{\today}
%
%
\begin{abstract}
A microscopic description of the quasielastic (\emph{p,n}) 
charge-exchange reaction (here, charge-exchange scattering 
between analogue states) is presented and discussed. 
Emphasis is focused on the spin-isospin structure of the projectile-target
coupling.
The model is a coupled-channel extension of the full-folding optical 
model approach (OMP) developed for nucleon elastic scattering, 
where emphasis is placed on retaining the genuine off-shell behavior of 
realistic effective interactions in the nuclear medium. 
The resulting non-local optical potentials are applied to the calculation
of (\emph{p,n}) differential cross sections, with particular emphasis on 
small-angle Fermi ($\Delta S=0$) cross-sections to isobaric analog states.
These parameter-free results provide a reasonable description of 
the $^{14}$C(\emph{p,n})-data at proton energies above $\sim$100 MeV,
but deteriorate for heavier targets. 
These shortcomings are analyzed and possible ways to correct them are
discussed.
\end{abstract}

\maketitle

\section{Introduction}
An issue of high current interest in nuclear research is the physics
behind systems and processes where the isovector component of the nuclear
interaction is relevant.
This particular aspect is of pivotal importance in understanding a wide
variety of phenomena such as extreme isospin asymmetric nuclear matter,
nuclear systems far from stability, collisions of
radioactive beams, the asymmetry energy of nuclear systems and (\emph{p,n})
charge-exchange processes \cite{Bao05,Dal05,Ron06,Rik02,Sch00,Noj96}.
A sound understanding of these phenomena may have significant impact on 
other areas of research such the physics of neutron stars, the formation 
of astrophysical objects, etc.
Our focus here is on the study of charge-exchange reactions at intermediate 
energies.
This effort attempts to provide a rigorous microscopic approach for the 
study of these reactions with emphasis on a detailed account of genuine 
energy, momentum and density effects in the nucleon-nucleon (\emph{NN}) 
effective interaction. 

The study of (\emph{p,n}) charge-exchange reactions at intermediate 
energies has received considerable attention during the past few decades. 
These reactions have been of great value in understanding the 
isovector modes of excitation of nuclei. 
At beam energies above 100 MeV, nucleon charge-exchange reactions can be 
considered as a one step process, thus allowing a rather clean separation 
of the nuclear structure from the underlying \emph{NN} effective interaction.
These arguments have resulted in the usual $t\rho$ structure of the 
coupling to the nucleus.
Furthermore, the weak strength of the isovector component of the interaction 
relative to its isoscalar counterpart has made suitable the use of the
distorted wave Born approximation (DWBA).
In spite of these advantageous considerations, no microscopic effort has been
able to satisfactorily describe the differential cross-section data without 
phenomenological adjustments \emph{a posteriori} of the nucleon-nucleus
(\emph{NA}) coupling.
This impediment is quite significant in that it points to 
a lack of understanding of the simplest process beyond elastic collisions.
Therefore, a comprehensive and consistent re-analysis of the problem is
relevant.

Studies of nucleon elastic scattering at intermediate energies 
(i.e. between 100 MeV and 1 GeV) have demonstrated the importance of
off-shell effects as accounted for within the full-folding OMP
for (\emph{NA}) scattering.
In particular, it was shown that the $t\rho$ scheme 
(as most often applied) is insufficient to satisfactorily describe 
intermediate energy data.
Furthermore, nuclear medium effects as included by using nuclear
matter $g$ matrices have provided significant improvements in the description
of the elastic scattering data \cite{Are02,Are89,Are95}.
The issue which naturally arises is whether these improvements within the
full-folding OMP to \emph{NA} elastic scattering are also 
important in charge-exchange processes.

Although significant advances have been made with the introduction 
of full-folding optical potentials in the description of
nucleon elastic scattering, none of these improvements have been 
included in charge-exchange reactions \cite{Amo00}. 
Furthermore, most of the reported (\emph{p,n}) reaction calculations, which
provide a reasonable description of the data, are semi-phenomenological
in the sense that well fitted potentials in the 
elastic channel are used to obtain the charge-exchange cross-section 
by means of the DWBA \cite{Che92,Cla84}. 
From this prospective additional theoretical effort is needed to clarify
the issues which prevent us from obtaining a unified understanding of 
these phenomena.
Along this line, the aim of the present work is to asses the importance
of treating the intrinsic off-shell behavior of, and medium corrections to,
realistic nuclear effective interactions in the calculation of 
nuclear charge-exchange reactions. 
For this purpose we present the spin-isospin formalism needed
to calculate these couplings within the non-local full-folding 
optical model and have developed the required coupled-channel scattering 
codes to solve the scattering problem exactly. 
Our work is focused on the $\Delta S=0$ isobaric analog state (IAS)
transitions for various targets and incident proton energies.

We organize this article in five sections as follows. 
In Section II we review the spin-isospin structure of the \emph{NA} 
coupling to display its formal structure suited for 
charge-exchange reactions from a finite nucleus.
In Section III we outline the coupled-channel approach for
studying the (\emph{p,n}) charge-exchange reactions with 
non-local potentials in the presence of the Coulomb interaction.
In Section IV we present and interpret results for representative
applications and in Section V we summarize the present work 
and draw the main conclusions. 
\section{Optical Potential: Spin and Isospin Considerations}

For simplicity in this discussion, we use the free $t$ matrix
as the \emph{NN} effective interaction and consider
nonrelativistic kinematics. Both of these restrictions will
be removed in the following sections.
Thus, the generalized optical potential for nucleon scattering and 
charge-exchange reactions may be written in momentum space as
\begin{eqnarray}
\langle \mathbf{k'} \nu\,' \mu' \mid U\mid {\bf k} \nu \mu \rangle 
&=&
\sum_{m,m',n,n'} \iint d{\bf p'} d{\bf p}
\langle F \mid 
\psi_{ \frac{1}{2}m',\frac{1}{2}n'}^{\dagger}({\bf p'}) \;
\overline\psi_{ \frac{1}{2}m,\frac{1}{2}n}({\bf p})\mid I\rangle \;  
\nonumber \\
&& 
\langle {\bf p}\,^{\prime }m^{\prime }\,n^{\prime },{\bf k'}\,
\nu\,^{\prime }\,\mu ^{\prime } \mid t\,\mid {\bf p}\,m\,n,{\bf k}\,\nu \,\mu
\rangle _{_{A}}.
\end{eqnarray}
Here $\langle t\rangle_{_{A}}$ is the antisymmetrized \emph{NN} $t$ matrix; 
$\nu $ and $\mu $ denote the initial spin and
isospin projections of the projectile and 
\[
\overline{\psi}_{\frac{1}{2}m,\frac{1}{2}n}({\bf p})=(-)^{\frac{1}{2}%
-m+\frac{1}{2}-n}\psi _{\frac{1}{2}-m,\frac{1}{2}-n}({\bf p})
\]
where $\psi _{\frac{1}{2}-m,\frac{1}{2}-n}({\bf p})$ annihilates a nucleon
with momentum ${\bf p}$ and spin and isospin projections $-m$ and $-n$,
respectively. 
The choice of the pair $(\mu ,\mu ^{\prime })$ is determined
by the reaction being considered. 
In the convention in which the proton has isospin projection $+\frac{1}{2}$, 
\[
(\mu ,\mu ^{\prime })=
( \textstyle\frac{1}{2} , \frac{1}{2} ) ,
( \frac{1}{2} ,-\frac{1}{2} ) ,
(-\frac{1}{2} , \frac{1}{2} ) ,
(-\frac{1}{2} ,-\frac{1}{2} ) \;
\]
for the (\emph{p,p}), (\emph{p,n}), 
        (\emph{n,p)}, and (\emph{n,n}) reactions, 
respectively. 
Following some recoupling to display the transferred quanta and the \emph{NN} 
spin and isospin, $U$ becomes: 
\begin{eqnarray}
U(S_{t}=0,T_{t}  = T_{p}) &=&
\frac{1}{4}\sum_{S_{^{p,}},T_{p}} \; 
\langle \nu\,^\prime\;\mid {\mathbb S}(S_{p},-\nu_{p})\mid \nu \rangle 
\langle \mu\,^\prime\;\mid {\mathbb T}(T_{p},-n_{t})\mid \mu \rangle 
(-)^{n_{t}}  
\nonumber \\
&&
\iint d{\bf p} \; d{\bf p'} 
\langle F \mid {\mathbb [\psi }_{\frac{1}{2}\frac{1}{2}}^{\dagger}({\bf p'})
\otimes
\overline{{\mathbb\psi}}_{\frac{1}{2}\frac{1}{2}}({\bf p}){]}^{00;T_{p}n_{t}}
\mid I\rangle \;
\nonumber \\
&&
\sum_{S,T,\,M,\,M^{\prime}} \;
\langle S\,S_{p}\,M\,\,-\nu _{p} 
\mid S\,M^{\prime }\rangle \,
(-)^{1-T}(2T+1)
\nonumber \\
&&
W( \textstyle{ \frac{1}{2} \frac{1}{2} \frac{1}{2} \frac{1}{2} } ;T_{p}T) 
\langle {\bf p}\,^{\prime }\,{\bf k'},S\,M^{\prime },T 
\mid t \, \mid {\bf p}\,{\bf k},S\,M,T\rangle _{A},
\end{eqnarray}
where 
\begin{equation}
\textstyle
{\lbrack\psi}_{\frac{1}{2}\frac{1}{2}}^{\dagger }({\bf p'})
\otimes 
\overline{\psi}_{\frac{1}{2}\frac{1}{2}}({\bf p}\,){]}^{SM;TN}
=\sum\limits_{m,m^{\prime },n,n^{\prime }}\langle 
\frac{1}{2} m^{\prime }%
\frac{1}{2}
m\mid SM\rangle \langle 
\frac{1}{2}
n^{\prime }%
\frac{1}{2}
n\mid TN\rangle \;\psi _{\frac{1}{2}m^{\prime },\frac{1}{2}n^{\prime
}}^{\dagger }({\bf p}^{\,\prime })\;\overline{{\mathbb \psi }}_{\frac{1}{2}m,%
\frac{1}{2}n}({\bf p}\,) \; ,  \label{crenil}
\end{equation}
defines the order of coupling. 
$T_{p}$ and $T_{t}$ denote the isospin transfers to the projectile and target, 
respectively. 
Similarly, $S_{p}$ and $S_{t}$ denote the corresponding spin transfers. 
Also, 
\[
{\mathbb S}(0,0)=1,\quad{\mathbb S}(1,-\nu _{p})=\sigma _{-\nu _{p}},\quad 
{\mathbb T%
}(0,0)=1,\quad {\mathbb T}(1,-n_{t})=\tau _{-n_{t}} \;,\quad 
\]
where $\sigma _{-\nu _{p}}$ ($\tau_{-n
_{t}}$) is the $-\nu _{p}$ ($-n_t$) spherical component of the
Pauli spin (isospin) matrix. 
The sum over $M$, $M^{\prime }$ projects out that part of the 
$t$ matrix which acts in the spin state $S$ and which is of rank $S_{p}$
in spin space. 
In particular, we write the $t$ matrix as a sum over the 
contributing ranks $(\lambda )$ in spin space 
\begin{equation}
t=\sum_{\lambda =0}^{2}\sum_{S,T}\;t_{ST}^{(\lambda )}\cdot O_{S}^{(\lambda
)}\;{\cal P}_{T} \; ,
\end{equation}
where 
\begin{equation}
O_{S}^{(\lambda )} = 
{\cal P}_{S} \; \delta _{\lambda 0} + (\vec{\sigma}_{1}+\vec{\sigma}_{2}) \; 
\delta _{S1} \; \delta _{\lambda 1} + (\vec{\sigma}_{1} \otimes 
\vec{\sigma}_{2})^{2} \; \delta _{S1} \; \delta _{\lambda 2} \; .
\end{equation}
Here, ${\cal P}_{S}$ and ${\cal P}_{T}$ are the projection operators onto the
\emph{NN} states of spin $S$ and isospin $T$, respectively. 
The optical potential $U$ corresponding to zero spin transfer to the 
target and $T_{t}=T_{p}$ isospin transfer to the target and projectile 
becomes: 
\begin{eqnarray}
U &=&
\sum_{S_{^{p,}},T_{p}}
\iint d{\bf p}\,d{\bf p'}
\rho_{FI}^{T_{t}}({\bf p}^{\,\prime },{\bf p})\cdot \;
\langle \mu\,' \mid {\mathbb T}(T_{p},\mu\,'-\mu)\mid \mu \rangle \;
\frac{
\langle \nu\,'\mid {\mathbb S}(S_{p},\nu\,'-\nu)\mid \nu \rangle
}{2S_{p}+1}\;
\nonumber \\
&&
\frac{1}{8}\sum\limits_{S\,T}\;(-)^{1-T}(2T+1)(2S+1)\,
W( \textstyle{ \frac{1}{2} \frac{1}{2} \frac{1}{2} \frac{1}{2}}
;T_{t}T)\;\langle {\bf p'}\,{\bf k'} 
\mid t_{ST}^{\,(S_{p})}\,\mid 
{\bf p}\,{\bf k}\rangle _{A} \;.  \label{Uopt}
\end{eqnarray}
The nuclear structure is contained in the mixed transition density defined by 
\begin{equation}
\rho_{FI}^{T_{t}}({\bf p'},{\bf p})=
2\;\langle F \mid 
{ [\psi}_{\frac{1}{2}\frac{1}{2}}^{\dagger }({\bf p'})
\otimes 
\overline{\psi}_{\frac{1}{2}\frac{1}{2}}({\bf p}){]}^{00;T_{t}}\mid
I\rangle \; ,   \label{TDST}
\end{equation}
which is an isoscalar or isovector as $T_{p}=T_{t}$ is $0$ or $1$. 
With this normalization for 
$\rho_{FI}^{T_{t}}$ , 
$\int d{\bf p}\;\rho _{II}^{0}({\bf p}^{\,},{\bf p})=A$ , 
the number of nucleons in state $I$. 
In the evaluation of the optical potential it is useful to make 
explicit use of the translational invariance characteristic of the 
free \emph{NN} interaction (or that in infinite nuclear matter) 
by writing it as 
\begin{equation}
\langle {\bf p'}\,{\bf k'} 
\mid t_{ST}^{\,(S_{p})}\,\mid 
{\bf p}\,{\bf k}\rangle _{A} = 
\delta ({\bf p}+\,{\bf k}-{\bf p'}-{\bf k'}) \;
\langle {\bm\kappa}^{\prime }
\mid t_{ST}^{\,(S_{p})}\,\mid 
{\bm\kappa}\rangle _{A} \;,  \label{2bt}
\end{equation}
where the relative momenta ${\bm\kappa}$ and ${\bm\kappa}^{\prime }$
are given by: 
\begin{equation}
{\bm\kappa}=\frac{{\bf p}-\,{\bf k}}{2} \;, \quad
{\bm\kappa}^{\prime}=\frac{{\bf p'}-\,{\bf k'}}{2} \;.  
\label{kappas}
\end{equation}
For brevity, it is also convenient to introduce the participating part of
the \emph{NN} interaction from Eq. (\ref{Uopt}) as 
\begin{equation}
\langle{\bm\kappa^{\prime}}\mid 
{\sf t}^{S_{p}T_{t}}\mid{\bm\kappa}\rangle _{A}=%
\frac{1}{8}\sum\limits_{S,T}(-)^{1+T}(2T+1)(2S+1) \; 
W( \textstyle{\frac{1}{2} \frac{1}{2} \frac{1}{2} \frac{1}{2}} ;T_{t}T)
\langle {\bm\kappa^{\prime}}\mid t_{ST}^{\,(S_{p})}\,\mid{\bm\kappa}%
\rangle _{A} \;,
\end{equation}
where W is the usual Racah coefficient, 
\[
W(\textstyle{ \frac{1}{2} \frac{1}{2} \frac{1}{2} \frac{1}{2}} ;T_{t}T) =
\left( 
\begin{array}{rr}
-\textstyle{\frac{1}{2}} & \textstyle{\frac{1}{2}} \\ 
 \rule[0pt]{0pt}{12pt}\textstyle{\frac{1}{2}} & \textstyle{\frac{1}{6}} 
\end{array}
\right)  \;,
\]
and $S_{t}=0$ for the optical potential.
%

\subsection{Scattering from a Fermi Gas ($N\geq Z$): an example}

For elastic nucleon scattering, the transition density 
of Eq. (\ref{TDST}) becomes 
\begin{equation}
\rho_{FI}^{T_{t}}({\bf p'},{\bf p})
=
2\;\delta ({\bf p'}-{\bf p}) \; 
\left[ 
\Theta (p_{F}(\pi )-p)+(-)^{T_{t}}\;\Theta (p_{F}(\nu )-p)
\right] \;,
\label{tdfg0}
\end{equation}
where $\Theta (x)$ is the Heaviside step function 
and $p_{F}(\pi )$ and $p_{F}(\nu )$
denote the Fermi momenta for protons and neutrons, respectively. 
The transition density appropriate for the excitation of the analogue state 
\emph{via} the charge-exchange reaction is 
\begin{equation}
\rho _{FI}^{T_{t}}({\bf p'},{\bf p})=2\sqrt{\frac{2}{N-Z}}%
\;\delta ({\bf p'}-{\bf p})\;[\Theta (p_{F}(\nu )-p)-\;\Theta
(p_{F}(\pi )-p)] \;,  
\label{tdfg1}
\end{equation}
where $\mid F\rangle $ $\,$satisfies 
\[
\mid F\rangle =\frac{1}{\sqrt{N-Z}} \; T_{+}\mid I\rangle \;.
\]
In the Fermi gas (FG) approximation we must have from 
Eqs. (\ref{2bt},\ref{kappas},\ref{tdfg0}),
${\bf p}={\bf p'},{\bf k}=\,{\bf k'}$ and therefore 
${\bm\kappa}={\bm\kappa}^{\prime }.$ 
This insures that only the $S_{p}=0$ part of the $t$ matrix 
participates as the spin-orbit and tensor terms do
not contribute to forward scattering in the \emph{NN} system. 
With these observations, the optical potential for elastic scattering
given by Eq. (\ref{Uopt}) becomes 
\begin{equation}
U_{FG} = 2\;\delta _{\mu ,\mu ^{\prime }} \; 
\delta _{\nu ,\nu\,^{\prime }} \; 
\delta ({\bf k}-\,{\bf k'})
\int d{\bf p} \; 
\left \{ 
\Theta \left [ p_{F}(\,l)-p \;\right ] 
\langle{\bm\kappa}\mid {\sf t}^{0}+{\sf t}^{1}\mid{\bm\kappa}\rangle_A \; +
\Theta \left [ p_{F}(u)-p) \;\right ] 
\langle{\bm\kappa}\mid {\sf t}^{0}-{\sf t}^{1}\mid{\bm\kappa}\rangle_A
\right \} \;,
\label{Upp}
\end{equation}
where $l$ ($u$) denotes target nucleons like (unlike) the projectile; 
e.g.  for proton scattering, $l$ corresponds to target protons 
and $u$ to target neutrons. 
Also, we have dropped the $S_{p}=0$ superscript on the reduced $t$ 
matrix ${\sf t}^{T_{t}}$.
Separating the isoscalar and isovector contributions to Eq. (\ref{Upp}) gives 
\begin{eqnarray}
U_{FG}=
2\;
\delta _{\mu ,\mu\,^{\prime }}\;
\delta _{\nu ,\nu\,^{\prime }}\;
\delta ({\bf k}-\,{\bf k'})\times  &&  \nonumber \\
\int d{\bf p}\;
\biglb (
\{\Theta [p_{F}(l)-p]+
  \Theta [p_{F}(u)-p]\}
\langle{\bm\kappa} 
& \mid & {\sf t}^{0}\mid{\bm\kappa}\rangle_{A}  
\nonumber \\
+ \{ 
    \Theta[p_{F}(l)-p] - 
    \Theta[p_{F}(u)-p] 
 \}
\langle{\bm\kappa} &\mid &
{\sf t}^{1}\mid{\bm\kappa}\rangle _{A}
\bigrb ) \;.
\end{eqnarray}
For the nucleon charge-exchange reaction exciting the isobaric analogue
state, the Fermi gas transition potential is

\begin{eqnarray}
\label{Upn}
U_{FG} &=&\frac{-4}{\sqrt{N-Z}}\;\delta _{\mu ^{\prime },\mu -1}\;\delta
_{\nu ,\nu\,^{\prime }}\;\delta ({\bf k}-\,{\bf k'})  \nonumber \\
&&\int d{\bf p}\;\{
 \Theta [p_{F}(\nu )-p]\;-
 \Theta [p_{F}(\pi )-p]\}
\langle {\bm\kappa}\mid {\sf t}^{1}\mid{\bm\kappa}\rangle _{A}.  
\end{eqnarray}

Some insight into the relationship of the optical model in the Fermi gas to
that in finite nuclei may be obtained by using the Fourier representation of
the momentum-conserving delta function appearing as a factor in $U_{FG}$
above, e.g. 
\[
\delta ({\bf k}-\,{\bf k'})=\frac{1}{(2\pi )^{3}}
\int d\,{\bf r}%
\;e^{i({\bf k}-\,{\bf k'})\cdot {\bf r}}, 
\]
together with the assumption that the $t$ matrix varies little 
over the range of $p$ allowed by the Fermi momentum. 
When this is done, the integrals over the step functions may be done giving 
\[ 
\frac{4\pi }{3}p_{F}^{3}=4\pi ^{3}\rho  \;,
\] 
for protons ($\pi$) and neutrons ($\nu$), and the optical potential 
for elastic scattering becomes 
\begin{equation}
\label{ufg1}
U_{FG}=
\delta _{\mu ,\mu\,^{\prime }}\;
\delta _{\nu ,\nu\,^{\prime }}\;\int d\,%
{\bf r}\;e^{i({\bf k}-\,{\bf k}^{\prime })\cdot {\bf r}}\;
( 
\;\rho_{l}\;
\langle{\bm\kappa}\mid {\sf t}^{0}+{\sf t}^{1}\mid{\bm\kappa}\rangle _{A}\;+
\rho_{u}\;
\langle{\bm\kappa}\mid {\sf t}^{0}-{\sf t}^{1}\mid{\bm\kappa}\rangle_{A}
)  \;,
\end{equation}
where ${\bm\kappa}$ is to be evaluated at $p=0$ , for example. 
In the strict Fermi gas model only the exponential above depends on ${\bf r}$ 
and we recover the delta function $\delta ({\bf k}-\,{\bf k'})$. 
When ${\bf k}=\,{\bf k'}$ the integration over ${\bf r}$ gives the volume 
$\Omega$ occupied by the gas and the product $\Omega\rho_{i}$
gives the number of target nucleons of type $i$. 
Therefore, 
\begin{equation}
U_{FG}({\bf k}=\,{\bf k'})=
\delta_{\mu ,\mu\,^{\prime }}\;
\delta_{\nu ,\nu\,^{\prime }}\;
(\;N_{l}
\langle{\bm\kappa}\mid {\sf t}^{0}+{\sf t}^{1}\mid{\bm\kappa}\rangle_{A}\;+
N_{u}
\langle{\bm\kappa}\mid {\sf t}^{0}-{\sf t}^{1}\mid{\bm\kappa}\rangle _{A}\;) \;,  
\label{ufg2}
\end{equation}
where $N_{l}$ ($N_{u}$) is the number of target nucleons like (unlike) the
projectile. 
Similarly, the non-vanishing diagonal elements of the optical potential 
for charge-exchange reactions is 
\begin{equation}
\label{ufgx}
U_{FG}({\bf k}=\,{\bf k'})=-2\sqrt{N-Z}\;
\delta _{\mu -1,\mu\,^{\prime }}\;
\delta _{\nu ,\nu\, ^{\prime }}\;
\langle{\bm\kappa}\mid {\sf t}^{1}\mid{\bm\kappa}\rangle _{A} \; .  
\end{equation}
If in Eq. (\ref{ufg1}), we use the local density approximation for 
$\rho^{[i]}$ by allowing it to be a function of ${\bf r}$, 
then 
\begin{equation}
\label{ufg3}
U_{FG}=
\delta _{\mu ,\mu\,^{\prime }}\;
\delta _{\nu ,\nu\,^{\prime }} \; 
(\;\tilde{\rho}_{l}(q)\:
\langle{\bm\kappa}\mid {\sf t}^{0}+{\sf t}^{1}\mid{\bm\kappa}\rangle _{A} \; + 
\tilde{\rho}_{u}(q)\:
\langle{\bm\kappa}\mid {\sf t}^{0}-{\sf t}^{1}\mid{\bm\kappa}\rangle _{A}\:)\;,
\end{equation}
where ${\bf q}={\bf k}-\,{\bf k'}$, the momentum transferred to the
target and $\tilde{\rho}_{i}$ is the Fourier transform of the density of
nucleons of type $i$. This is an approximate form of the usual $t\rho $
approximation to the optical potential.
%

\subsection{Scattering From Finite Nuclei}

Results for finite nuclei may be obtained by expanding the creation and
annihilation operators of Eq. (\ref{crenil}) in a (spherical) shell-model
basis. 
The result for arbitrary (allowed) spin and isospin transfer is 
\begin{eqnarray}
\left [  
\psi_{\frac{1}{2}\frac{1}{2}}^{\dagger }({\bf p'})
\otimes 
\overline{\psi}_{\frac{1}{2}\frac{1}{2}}({\bf p}) 
\right ]^{SM;TN}
& = &
\sum_{nlj,n^{\prime }l^{\prime }j^{\prime }}\;
u_{nlj}(p)\;u_{n^{\prime }l^{\prime }j^{\prime }}(p^{\prime })  
\sum_{L,J}\;
[ \hat{L}\;\hat{J}\hat{\jmath}\;\hat{\jmath}^{\prime}\;
(-)^{l^{\prime }+L+S+J}]
\left( 
\begin{array}{lll}
l^{\prime } & \frac{1}{2} & j^{\prime } \\ 
l & \frac{1}{2} & j \\ 
L & S & J
\end{array}
\right) \times
\nonumber \\
\{[Y_{l^{\prime }}({\bf\hat p}^{\prime })\otimes Y_{l}({\bf\hat p}%
)]^{L}
&\otimes&
[c_{n^{\prime }l^{\prime }j^{\prime };\frac{1}{2}}^{\dagger
}\otimes \bar{c}_{n^{\prime }l^{\prime }j^{\prime };\frac{1}{2}%
}]^{J;TN}\;\}^{SM} \;,  
\label{crenil2}
\end{eqnarray}
where, in this equation, $L,S,J,T$ denote transferred quanta $L_{t},T_{t}$ etc. 
Also, $c^{\dagger }$ and $\bar{c}$ denote creation operators
for particles and holes, respectively, and $u_{nlj}(p)$ is the
radial part of a shell-model orbital in momentum space. 
The symbol in rounded parentheses is a nine-j symbol and $\hat{x}$ 
denotes $\sqrt{2x+1}$.
Inserting Eq. (\ref{crenil2}) into Eq. (\ref{TDST}) gives the transition
density for a finite nucleus 
\begin{eqnarray}
\rho _{FI}^{S_{t}T_{t}}({\bf p'},{\bf p})
&=&
2\;\sum_{nlj,n^{\prime }l^{\prime }j^{\prime }}\;
u_{nlj}(p)\;u_{n^{\prime}l^{\prime }j^{\prime }}(p^{\prime })\;
\sum_{L_{t}\;J_{t}}\;
\left[
\hat{L}_{t}\;\hat{J}_{t}\text{\ }\hat{\jmath}\;\hat{\jmath}^{\prime }\;
(-)^{l^{\prime}+L_{t}+S_{t}+J_{t}}
\right]
\left( 
\begin{array}{lll}
l^{\prime } & \frac{1}{2} & j^{\prime } \\ 
l & \frac{1}{2} & j \\ 
L_{t} & S_{t} & J_{t}
\end{array}
\right) \times
\nonumber \\
&\{&
[
Y_{l^{\prime }}({\bf\hat p}^{\prime })\otimes Y_{l}({\bf\hat p}%
)]^{L_{t}}\otimes \langle F\mid [c_{n^{\prime }l^{\prime }j^{\prime };\frac{1%
}{2}}^{\dagger }\otimes \bar{c}_{nlj;\frac{1}{2}}]^{J_{t};T_{t}N_{t}}\;\mid
I \, \rangle \; \}^{S_{t}\,M_{t}} \; . \label{tdst2}
\end{eqnarray}
Although only the $S_{t}=0$ part of the above density is required for the
usual case of calculating the optical potential, it is useful to have the
more general result for later use. 
For the special case in which the transferred 
quanta $(L_{t}\,S_{t}\,J_{t})$ are 
either assumed or restricted to be zero, Eq. (\ref{tdst2}) becomes 
\begin{eqnarray}
\rho _{FI}^{T_{t}}({\bf p'},{\bf p}) &=&\frac{-\sqrt{2}}{4\pi }%
\sum_{n\,l\,j}\;u_{nlj}(p)\;u_{nlj}(p^{\prime })\;P_{l}
({\bf\hat p}^{\prime}\cdot {\bf\hat p})  \nonumber \\
&&\sum_{\mu \,\nu \,\nu\,^{\prime }}\;(-)^{\frac{1}{2}-\nu }
\langle \textstyle{ \frac{1}{2} \frac{1}{2} } \,\nu\,^{\prime }\,\nu 
\mid T_{t}\,N_{t}\,\,\rangle \;
\langle F\mid c_{n^{\prime }lj\,-\mu ;\frac{1}{2}\nu\,^{\prime }}^{\dagger }\;
c_{nlj\,-\mu ; \frac{1}{2}\,-\nu }\;\mid I\rangle \; . \label{tdst3}
\end{eqnarray}

For the special case of elastic scattering in which 
$\mid F\;\rangle =\mid I\;\rangle$, 
$n=n^{\prime }$ and $N=0$ , the mixed density reduces to 
\begin{equation}
\rho _{FI}^{T_{t}=1}({\bf p'},{\bf p})=\frac{1}{4\pi }%
\sum_{n\,l\,j}\;u_{nlj}(p)\;u_{nlj}(p^{\prime})\;
\left [
Z_{nlj}+(-)^{T_{t}}\,N_{nlj}
\right ] \;
P_{l}({\bf\hat p}^{\prime }\cdot{\bf\hat p}) \;,
\label{gs0}
\end{equation}
where $Z_{nlj}$ and $\,N_{nlj}$ denote the number of protons and neutrons in
the state $(nlj)$ where it has been assumed that the orbitals for neutrons
and protons are the same. 
For the (\emph{p,n}) charge-exchange reaction to the isobaric analogue state, 
$\nu =\nu\,^{\prime}=\frac{1}{2}\;,\;T_{t}=N_{t}=1\;$
and the Clebsch-Gordan coefficient is unity giving 
\begin{equation}
\rho _{FI}^{T_{t}=1}({\bf p'},{\bf p})=-\sqrt{\frac{2}{N-Z}}\frac{%
1}{4\pi }\sum_{n\,l\,j}\;u_{nlj}(p)\;u_{nlj}(p^{\prime
})\;[\,N_{nlj}-Z_{nlj}]\;P_{l}({\bf\hat p}^{\prime }\cdot{\bf\hat p}).  
\label{ias}
\end{equation}

From Eq. (\ref{Uopt}), the optical potential for elastic scattering
associated with the transition density of Eq. (\ref{gs0}) reduces to
\begin{eqnarray}
U \to U_{pp,nn}
&=&
\delta _{\mu \mu ^{\prime }}\frac{1}{4\pi }\sum_{n\,l\,j}\;\int d\,{\bf p}%
\;u_{nlj}(p)\;u_{nlj}(p^{\prime }) \;
P_{l}({\bf\hat p}^{\prime}\cdot {\bf\hat p}) \, %
\sum_{S_{p}=0,1}\frac{\langle \nu\,^{\prime }\mid 
{\mathbb S}(S_{p},\nu\,^\prime-\nu)\mid \nu
\rangle }{2S_{p}+1} \times \nonumber \\
 & &\left( N_{nlj}^{\,l}\,\langle{\bm\kappa}^{\prime}
\mid {\sf t}^{S_{p}0}+{\sf t}%
^{S_{p}1}\mid{\bm\kappa}\rangle _{A}\;+N_{nlj}^{\,u}\,
\langle{\bm\kappa}^{\prime}%
\mid {\sf t}^{S_{p}0}-{\sf t}^{S_{p}1}\mid{\bm\kappa}\rangle _{A}
\right ) \;,
\label{Ufn}
\end{eqnarray}
where $N_{nlj}^{\,l}$ and $N_{nlj}^{\,u}$ denote the number of target
nucleons in the state $(nlj)$ which are like and unlike (isospin projection)
the projectile and ${\bf p'}={\bf p}+{\bf q}$. 
Similarly, the optical potential mediating the (\emph{p,n}) reaction to
the isobaric analogue of the target ground state is
\begin{eqnarray}
U \to U_{x} 
&=&-\delta _{\mu -1,\mu ^{\prime }}\frac{2}{\sqrt{N-Z}}\frac{1}{4\pi }%
\sum_{n\,l\,j}\;\int d\,{\bf p}\;u_{nlj}(p)\;u_{nlj}(p^{\prime}) \;
( \,N_{nlj}-Z_{nlj} ) \; P_{l}(\hat{\bm p}^{\prime }\cdot \hat{\bm p}) \nonumber \\
 & &\sum_{S_{p}=0,1} \; \frac{\langle\nu\,^{\prime }\mid
{\mathbb S}(S_{p},\nu\,^\prime-\nu)\mid \nu
\rangle }{2S_{p}+1}\cdot 
\langle{\bm\kappa}^{\prime}\mid {\sf t}^{S_{p}1}\mid
{\bm\kappa}\rangle _{A}\;.  \label{Ufnx}
\end{eqnarray}

The above expressions can be further simplified in the case of non-spin
transfer to the projectile. 
Denoting the proton and neutron mixed densities by $\rho_p$ and $\rho_n$,
\begin{eqnarray}
\label{mix_z}
\rho_{p}({\bf p'}, {\bf p} ) &=& 
\sum_{n\,l\,j} 
\;u_{nlj}(p)\;u_{nlj}(p^{\prime}) \;
Z_{nlj} \; P_{\,l}(\hat{\bm p}^{\prime }\cdot \hat{\bm p}) \;, \nonumber \\
\rho_{n}({\bf p'}, {\bf p} ) &=& 
\sum_{n\,l\,j} 
\;u_{nlj}(p)\;u_{nlj}(p^{\prime}) \;
N_{nlj} \; P_{\,l}(\hat{\bm p}^{\prime }\cdot \hat{\bm p})\; ,
\end{eqnarray}
we are left with the following three terms to evaluate
\begin{eqnarray}
\label{ff_pn}
U_{pp}
&=&
\frac{1}{4\pi }\int d\,{\bf p}  
 \left[
\rho_{p}({\bf p'}, {\bf p} )
 \langle{\bm\kappa}^{\prime}\mid 
 {\bf t}^{0}+{\sf t}^{1}
 \mid{\bm\kappa}\rangle _{A} +
\rho_{n}({\bf p'}, {\bf p} )
 \langle{\bm\kappa}^{\prime}\mid 
 {\bf t}^{0}-{\bf t}^{1}
 \mid{\bm\kappa}\rangle _{A}
 \right ] \;, \nonumber \\
U_{x}
&=&
\frac{-1}{2\pi\sqrt{N-Z}}\int d\,{\bf p}  
 \left[
 \rho_{n}({\bf p'}, {\bf p} )- \rho_{p}({\bf p'}, {\bf p} )
 \right]
 \langle{\bm\kappa}^{\prime}\mid 
 {\sf t}^{1}
 \mid{\bm\kappa}\rangle _{A} \;,
 \nonumber \\
U_{nn}
&=&
\frac{1}{4\pi }\int d\,{\bf p}  
 \left[
\rho_{p}({\bf p'}, {\bf p} )
 \langle{\bm\kappa}^{\prime}\mid 
 {\bf t}^{0}-{\sf t}^{1}
 \mid{\bm\kappa}\rangle _{A} +
\rho_{n}({\bf p'}, {\bf p} )
 \langle{\bm\kappa}\mid 
 {\bf t}^{0}+{\bf t}^{1}
 \mid{\bm\kappa}\rangle _{A}
 \right ] \; .
\end{eqnarray}

\subsection{{\it \bfseries In-medium} calculations}
The optical potential $U_{pp}$, $U_{nn}$ and $U_{x}$ obtained in the 
previous section requires the convolution of single-particle 
wavefunctions with a two-body effective interaction.
Simpler expressions for these potentials are obtained with the use
of the Slater approximation to the ground-state mixed density in the
case of spin-saturated targets.
This approximation was examined in Ref. \cite{Are90b} in the context
of nucleon elastic scattering at intermediate energies, and its effects
are reported to be only noticeable at the larger scattering angles.
Furthermore, this representation of the mixed density yields  
a simple prescription for including explicit medium effects by means
of infinite nuclear matter $g$ matrices. 
Following the same arguments discussed in Ref. \cite{Are95} to 
incorporate nuclear medium effects, the first of the potentials expressed
in Eqs. (\ref{ff_pn}) takes the general form (c.f. Eq. (21) of 
Ref. \cite{Are02})
\begin{equation}
\label{upp}
U_{pp} ( {\bm k'}, {\bm k}) =
\int d\;{\bm R} \;
e^{i({\bm k}' -{\bm k})\cdot {\bm R}} \left [
\rho_p({\bm R})\; \bar g_{pp}({\bm k}',{\bm k})  +
\rho_n({\bm R})\; \bar g_{np}({\bm k}',{\bm k})
\right ] \; ,
\end{equation}
where $\rho_p$ and $\rho_n$ are the local proton and neutron point
densities, respectively, and $\bar g_{NN}$ represent off-shell
Fermi-averaged amplitudes in the appropriate \emph{NN} channel.
More explicitly, in the context of infinite nuclear matter these 
density-dependent amplitudes are given by
\begin{equation}
\label{g_on}
\bar g_{NN}({\bm k}',{\bm k}) =
\frac{3}{4\pi \hat k^3} \int
\Theta(\hat k - |{\bm P}| )
g_{{\bm K}+ {\bm P}}({\bm\kappa}',{\bm\kappa};\:\sqrt{s};\:\bar\rho\:)
\;d{\bm P}\; ,
\end{equation}
where $g_{{\bm K}+ {\bm P}}({\bm\kappa}',{\bm\kappa};\sqrt{s};\bar\rho)$
is taken as the $g$ matrix for symmetric nuclear matter of
density $\bar\rho$.
Here, 
\begin{equation}
{\bm K}= \textstyle{\frac{1}{2}} ({\bm k}+{\bm k}')\;, \quad
{\bm P}= \textstyle{\frac{1}{2}} ({\bm p}+{\bm p}')\;, \quad
{\bm k}+{\bm p}={\bm k}'+{\bm p}'\;,
\end{equation}
and the relativistically corrected relative momenta 
${\bm\kappa}$ and ${\bm\kappa}^\prime$ are given by
\begin{equation}
\label{rel_mtum}
{\bm\kappa}        = W {\bf k} - (1-W){\bf p}\; , \qquad
{\bm\kappa}^\prime = W^\prime\: {\bf k}^\prime-(1-W^\prime)\:{\bf p}^\prime\; ,
\end{equation}
with $W$ and $W^\prime$ scalar functions of the momenta of the 
colliding particles with relativistic kinematics built in \cite{Are02}.
Additionally, the local momentum $\hat k$ is taken from
$\hat k^3=3\pi^2\bar\rho/2$, with $\bar\rho\to[\rho_p(R)+\rho_n(R)]/2$, 
the isoscalar local density at radius $R$.
Analogous expressions are obtained for $U_{x}$ and $U_{nn}$.

\section{Coupled-channel calculations}

Our study focuses on quasielastic scattering to the isobaric analog state.
Considering explicitly the isospin degrees of freedom of the scattering
waves in the form of outgoing proton and neutron wavefunctions, we obtain a
non-local version of the coupled-channel \cite{Lan62,Hof73} equations
\begin{equation}
\label{lane1}
\left [
{\begin{array}{cc} \hat K_p + U_{pp}^{(s)}+V_{C} &  U_{x} \\
                   U_{x}  & \hat K_n + U_{nn}
\end{array}}
\right ]
\left(
{\begin{array}{c} \Psi_p \\ \Psi_n \end{array}}
\right)
= \left (
\begin{array}{c}
E_p \Psi_p \\
E_n \Psi_n
\end{array}
\right) \;.
\end{equation}
Here $U_{nn}$ and $U_{x}$ are the non-local potentials
given by Eqs. (\ref{Ufn}) and (\ref{Ufnx}) in the previous Section.
Furthermore, $U_{pp}^{(s)}$ represents the short-range coupling
between the charged projectile and the target protons, i.e. 
the hadronic-plus-Coulomb contribution 
with the point-Coulomb potential, $V_C=Ze^2/r$, subtracted.
A formal solution to these equations is given in the form of the
Lippmann-Schwinger integral equation with proton waves $\Phi_{p}$ 
in the entrance channel
\begin{equation}
\label{lane2}
\left(
\begin{array}{c}
{\Psi _{p}} \\
\Psi _{n}%
\end{array}%
\right) =\left(
\begin{array}{c}
{\Phi_{p}} \\
0%
\end{array}%
\right) +\left[
\begin{array}{ll}
{{\cal G}_{p}U_{pp}^{(s)}} & {\cal G}_{p}U_{x} \\
 {\cal G}_{n}U_{x}         & {\cal G}_{n}U_{nn}%
\end{array}%
\right] \left(
\begin{array}{c}
{\Psi _{p}} \\
\Psi _{n}%
\end{array}%
\right) \;,  
\end{equation}%
where ${\cal G}_{p}$ and ${\cal G}_{n}$ are the Green's functions for
outgoing protons and neutrons, respectively.
In coordinate space, for partial wave $l$, these propagators are 
expressed in terms of the Coulomb (and Bessel) spherical waves
\begin{eqnarray}
\label{propag}
{\cal G}_l^{(+)}(r,r';k) &=& -\frac{i}{k}\,\frac{2\bar\epsilon}{\hbar^2} \:
F_l(\eta;kr_{<}) \left [ F_l(\eta;kr_{>}) + i G_l(\eta;kr_{>}) \right ] 
\nonumber \\
&\equiv& 
-\frac{i}{k}\,\frac{2\bar\epsilon}{\hbar^2} \: 
F_l(\eta;kr_{<}) H_l^{(+)}(\eta;kr_{>})\; .
\end{eqnarray}
Here $\eta=Ze^2/\hbar v$, 
represents the charge parameter for the \emph{NA} coupling, 
with $v$ the corresponding relative velocity, 
and $\bar\epsilon$ the ejectile-residual nucleus reduced energy.
For uncharged particles $F_l$ and $G_l$ become the usual 
Ricatti-Bessel functions with the following the phase conventions:
${F}_l(0;t) = t j_l(t)$; $j_0(t)=\sin t/t$, and
${G}_l(0;t) = t n_l(t)$; $n_0(t)=-\cos t/t$.

The primary input in these equations is the optical potential which 
we obtain following a current version of the full-folding OMP
to nucleon scattering \cite{Are02,Are95}. 
In this scheme, the optical potential is calculated in momentum space
leading to a general non-local structure in coordinate space, a feature
which is retained throughout.
An exact treatment of the Coulomb interaction in the presence 
of this non-local coupling is performed in coordinate space by
solving Eq. (\ref{lane2}) once the Fourier transform of the hadronic
contribution have been performed.
Thus, the scattering amplitude is readily obtained from the asymptotic
behavior of the radial scattering waves from Eq. (\ref{lane2}) which, 
schematically, has the form
\begin{equation}
\label{asymp}
\left[
\begin{array}{l}
u_{p,l}(r) \\
u_{n,l}(r) 
\end{array}
\right]
\sim
\left[
\begin{array}{c}
\frac{1}{k_p}F_l(\eta;k_pr) \\
0
\end{array}
\right]
+
\left[
\begin{array}{r}
{\Delta}_{pp}\:\frac{1}{k_p}\: {H}_l^{(+)}(\eta;k_pr) \\
{\Delta}_{pn}\:\frac{1}{k_n}\: {H}_l^{(+)}(0;k_nr) 
\end{array}
\right] \;,
\end{equation}
with $k_p$ and $k_n$ the momenta of the outgoing protons 
and neutrons, respectively. 
The above result provides alternative forms for obtaining the scattering
amplitude, namely by simple identification of the asymptotic
form of the solution, or by explicit evaluation of the matrix elements,
i.e.
\begin{eqnarray}
\label{amplit}
{\Delta}_{pp}^{( l )} &=&
\langle\; 
U_{pp}^{(s)}\:u_{p,l} + U_{x}\:u_{n,l}
\;\rangle 
\equiv ik_p\,f_{pp}^{(l)}\;,
\nonumber\\
{\Delta}_{pn}^{( l )} &=&
\langle\; 
U_{x}\:u_{p,l} + U_{nn}\:u_{n,l}
\;\rangle 
\equiv ik_n\,f_{pn}^{(l)}\;.
\end{eqnarray}
More explicitly,
\begin{equation}
\label{intamplit}
\langle\;U_{pp}^{(s)}\:u_{p,l}\;\rangle= 
-i\frac{2\bar\epsilon_p}{\hbar^2}\:
\int_{0}^{\infty}dr
\int_{0}^{\infty}dr'\:
F_l(\eta;k_pr')\:[\,r'U_{pp}^{(s)}(r',r)\;r\,]\:\:u_{p,l}(r)\;.
\end{equation}
Analogous expressions hold for 
$\langle\;U_{pn}\:u_{n,l}\;\rangle$ and
$\langle\;U_{nn}\:u_{n,l}\;\rangle$.

The results obtained with this procedure were compared with 
DWBA results and showed almost unnoticeable differences. 
\section{Applications}

The scattering calculations performed in this work differ from 
most reported applications of quasielastic nucleon
exchange processes to date. 
Thus, it is important to spell out the differences explicitly.
The optical potential is obtained \emph{via} a convolution of the
\emph{in-medium} nuclear matter $g$ matrix with the target ground state
mixed density. We have considered six different \emph{NN} potential models
in the construction of the corresponding $g$ matrix, i.e. 
the Paris potential \cite{Lac80},
the Nijmegen I, Nijmegen II and Reid 93 potentials \cite{Sto94}
the Argonne AV18 potential \cite{Wir95}
and the charge-dependent Bonn potential \cite{Mac01}.
The $g$ matrix is calculated in momentum space fully off shell, 
evaluated at eight different densities and nearly twenty momentum pairs, 
for all allowed \emph{NN} states up to $J=7$, 
in a square mesh of relative momenta up to 15 fm$^{-1}$. 
The Fermi motion integrals involving the $g$ matrix are made following
Ref. \cite{Are02}, where relativistic kinematics at the \emph{NN} level 
are included in the folding integrals, supplemented by a proper
treatment of the deuteron pole contribution \cite{Are94}.
For simplicity, and in order to keep our discussion focused
within a single scheme, we have restricted the use of the \emph{NN} 
charge-dependent potential models to their \emph{NN} $T_z=0$ components.

The results reported in this study are based on semi-phenomenological
densities. Proton densities were obtained from parametrized charge
densities \cite{Jag74}, unfolding the electromagnetic proton size to
obtain the corresponding point density.
Neutron densities were obtained by adding to the proton density
a harmonic oscillator orbital of the form 
$\psi\sim \,r^l\,e^{-a_n^2r^2/2}$.
In the case of $^{90}$Zr, a 3pG charge density was used,
with the neutron excess characterized by $a_n$=0.472 fm$^{-1}$.
This construction yields a proton (neutron) root-mean-square radius
($R_{RMS}$) of 4.198 fm (4.363 fm).
A similar construction was used for the $^{48}$Ca densities, with
a 3pF charge density for the closed shells with the neutron excess 
characterized by $a_n$=0.524 fm$^{-1}$.
The resulting proton (neutron) $R_{RMS}$ is  3.374 fm (3.580 fm).
In the case of $^{14}$C, the proton density was represented by a
modified harmonic oscillator.
The neutron excess was characterized with $a_n$=0.510 fm$^{-1}$.
The proton (neutron) $R_{RMS}$ is in this case 2.427 fm  (2.611 fm).
The scattering results obtained with these densities were consistent
with a variety of choices for the neutron-excess wavefunctions.

\subsection{Results for $^{14}$C(\emph{p,n})}

\begin{figure}[ht]
\includegraphics[scale=0.50,angle=00]{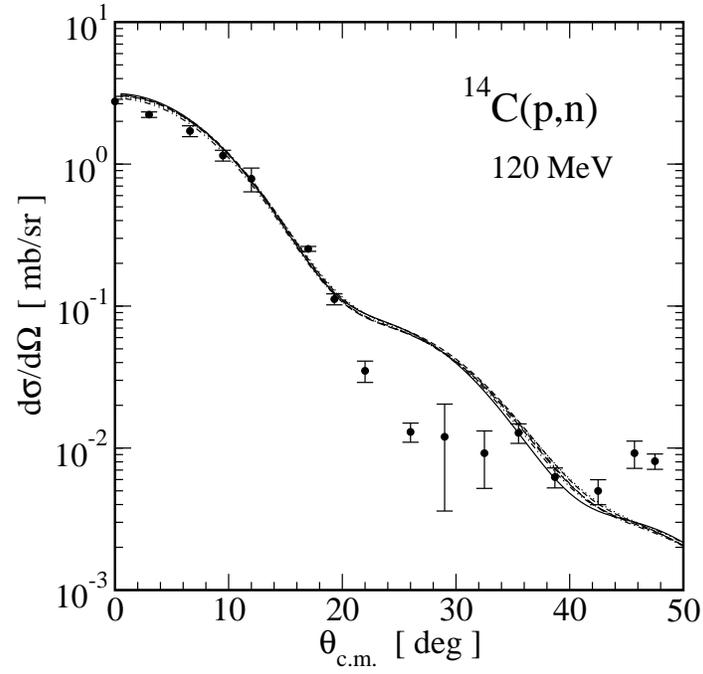} 
\medskip
\caption{\protect\small
\label{pnC14_120}
         Measured  \cite{Rap83} and calculated differential cross-section 
	 for $^{14}$C($p,n$) at 120 MeV.
	 For reference to the curves see the text.
        }
\end{figure}

In Fig. \ref{pnC14_120} we present the measured \cite{Rap83} and
calculated differential cross-section
for the quasielastic $^{14}$C(\emph{p,n}) reaction following the $g$-matrix
approach described in the previous section.
The excitation energy to the IAS, $E_x$, is 2.31 MeV \cite{And91}. 

\begin{figure}
\includegraphics[scale=0.90,angle=00] {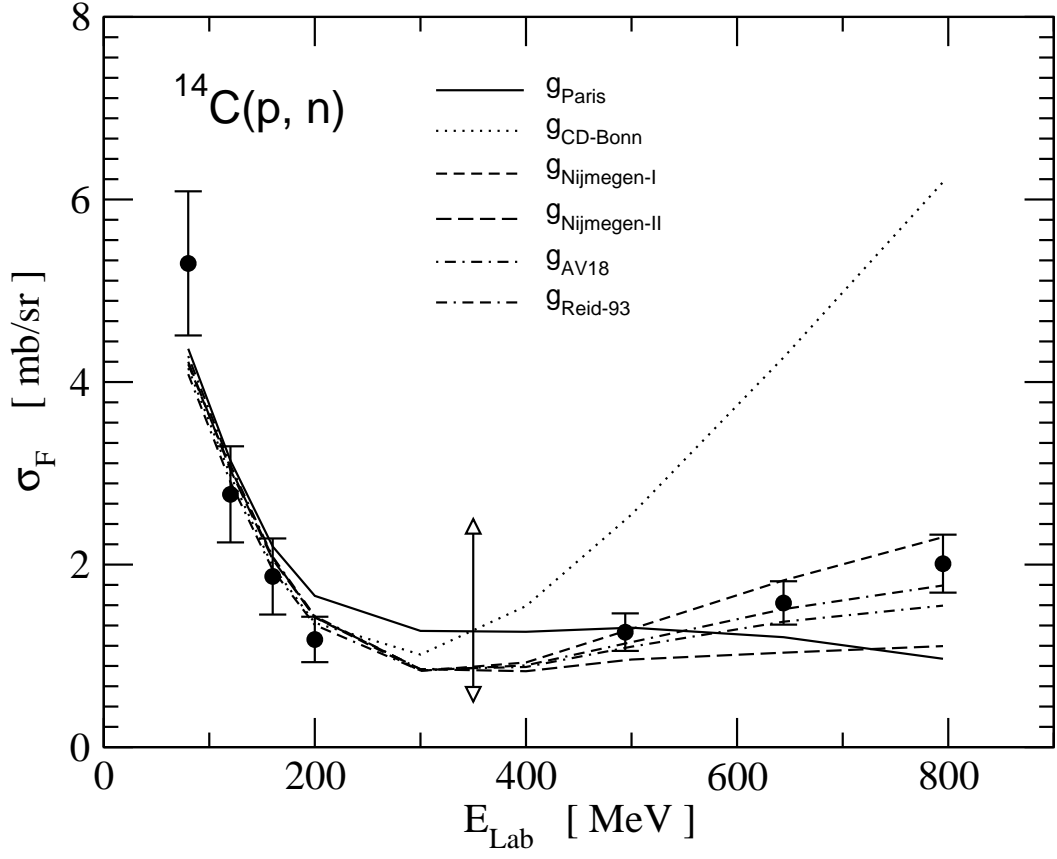} 
\medskip
\caption{\protect\small
\label{fermiDAT}
   The zero-degree Fermi cross section based on $g$-matrix full-folding
   optical model potentials using different \emph{NN} potential models. 
   For reference to the data see text.
 The dashed line at 350 MeV indicates the energy from which the
 calculations should be taken as extrapolations of all \emph{NN} models. 
        }
\end{figure}

In this figure, and also in Figs. 
\ref{pnC14_120}, \ref{fermiDAT} and \ref{pnCa48_135}-\ref{pnZr90_160},
we show results for the 
Paris (solid curves),
CD-Bonn (dotted curves),
Nijmegen I (short-dashed curves),
Nijmegen II (long-dashed curves),
Argonne AV18 (dot-dashed curves), and
Reid-93 (dot-dot-dashed curves) \emph{NN} potential models.
Although in some cases the overlapping of the curves prevents 
their identification, this comparison allows one to visualize 
a moderate sensitivity of the calculated quantities to the choice of 
the \emph{NN} model.
Such is the case of  Fig. \ref{pnC14_120}, where the description of
the differential cross section for scattering angles less than 25 degrees
is reasonably well described by all \emph{NN} potential models considered.
A departure from the data is observed at larger angles,
suggesting missing effects.

Another quantity of particular interest is 
the so-called Fermi cross section ($\sigma_F$), that is to say, 
the forward-angle (zero-degree)
differential cross section of the quasielastic (\emph{p,n}) reaction.
In Fig. \ref{fermiDAT} we present the measured and calculated 
Fermi cross sections for $^{14}$C(\emph{p,n}).
The data are from Refs. \cite{Tad82,Sug90}, and the various curves
correspond to the six \emph{NN} models considered.
A note of caution should be kept in mind before interpreting this figure.
Although the $g$-matrix full-folding optical model calculations
account for relativistic kinematics,
all the \emph{NN} potential models have been developed for energies 
below pion production threshold.
Thus, the results in Fig. \ref{fermiDAT} above 350 MeV represent
extrapolations of the \emph{NN} models. 

Below 150 MeV all \emph{NN} models provide reasonably good descriptions 
of $\sigma_F$, albeit all of them underestimate the 80-MeV datum. 
As the energy increases, beginning at $\sim$150 MeV results using 
the Paris potential depart from the others, being followed by those using
the CD-Bonn potential near 300 MeV.
Above pion production threshold the CD-Bonn potential exhibits 
a distinctive uniform growth, being followed less strongly by 
the Nijmegen I and Reid-93 models.
Over the whole energy range both the Argonne AV18 and Nijmegen II potentials
yield results in closest agreement with the data.
In contrast, the Paris potential exhibits qualitative differences from 
the other models, particularly the depth of its minimum near 300 MeV.

\begin{figure}
\includegraphics[scale=0.60,angle=00]{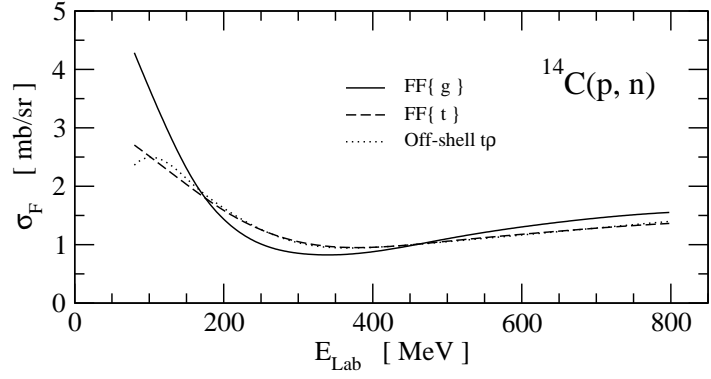} 
\medskip
\caption{\protect\small
\label{fermiGvsT}
 The zero-degree Fermi cross section based on full-folding 
 OMP using the $g$ matrix (solid curve) 
 and $t$ matrix (dashed curve). 
 Off-shell $t\rho$ results are represented with the dotted curve.
 All three results use the Argonne AV18 \emph{NN} potential.
        }
\end{figure}

In a broad sense, the role of medium effects have been subject 
of significant interest in the description of \emph{NA} processes.
In Fig. \ref{fermiGvsT} we contrast results for the Fermi cross
section as function of the proton incident energy using the $g$ matrix
(solid curve) and $t$ matrix (dashed curve).
Here we also include results based on the off-shell $t\rho$ approximation
(dotted curve),
a limiting case of the full-folding approach using the free $t$ matrix
and suppressing Fermi motion effects in the \emph{NN} interaction.
In all cases the Argonne AV18 potential is used. 
We observe that the differences between the $g$-
and $t$-matrix approaches become more sizable at the lower energies. 
In general, the difference between the $g-$ and $t-$matrix approaches 
above 150 MeV does not exceed 0.2 mb/sr, a useful upper bound of medium 
effects in (\emph{p,n}) charge-exchange processes.
At energies below 150 MeV the Fermi cross section exhibit sizable 
differences between both approaches.

\subsection{Results for $^{48}$Ca(\emph{p,n}) and $^{90}$Zr(\emph{p,n})}

\begin{figure}[ht]
\includegraphics[scale=0.50,angle=00]{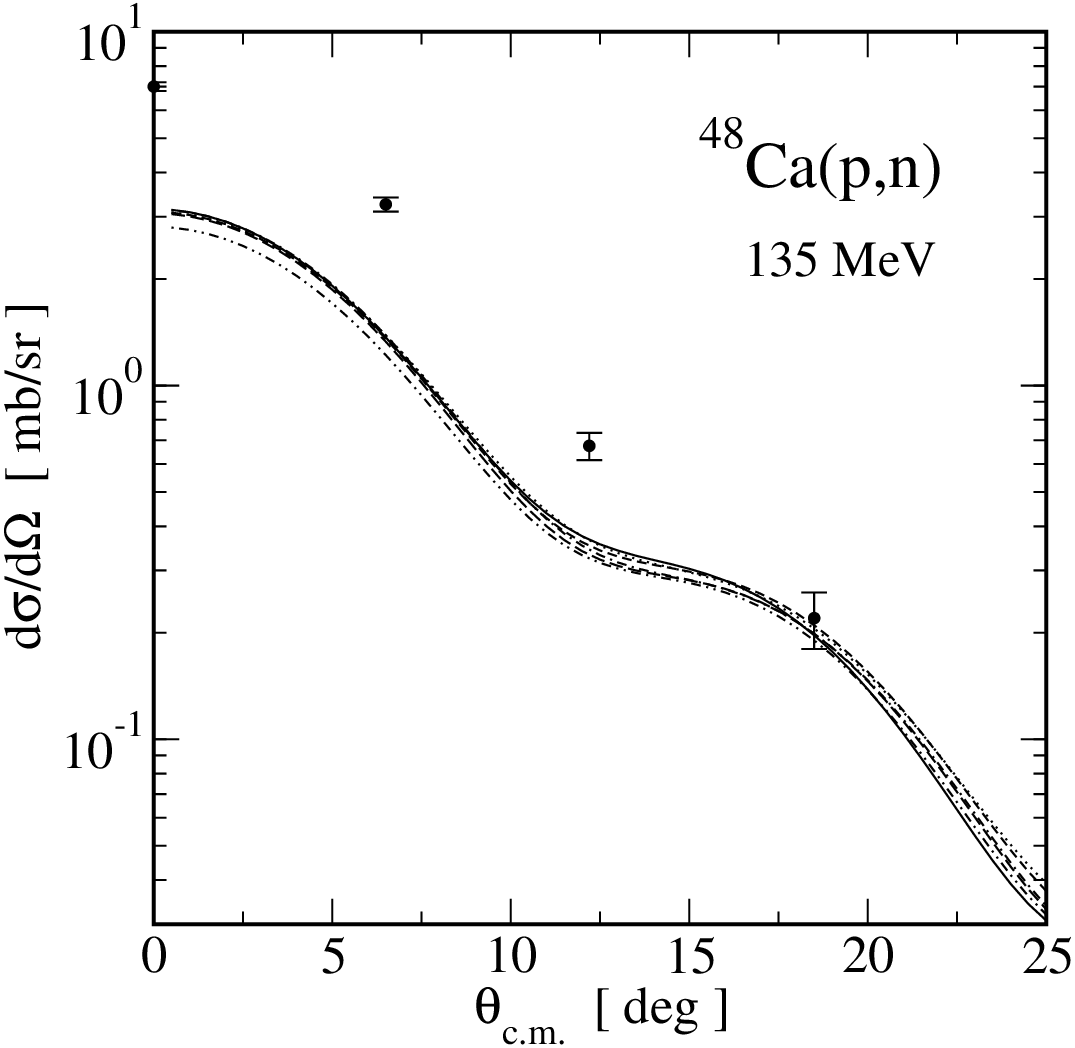} 
\medskip
\caption{\protect\small
\label{pnCa48_135}
         Measured \cite{And91} and calculated differential cross-section 
	 for $^{48}$Ca($p,n$) at 135 MeV.
	 The curve patterns follow the same convention as in
	 Fig. \ref{fermiDAT}. 
        }
\end{figure}

\begin{figure}[ht]
\includegraphics[scale=0.50,angle=00] {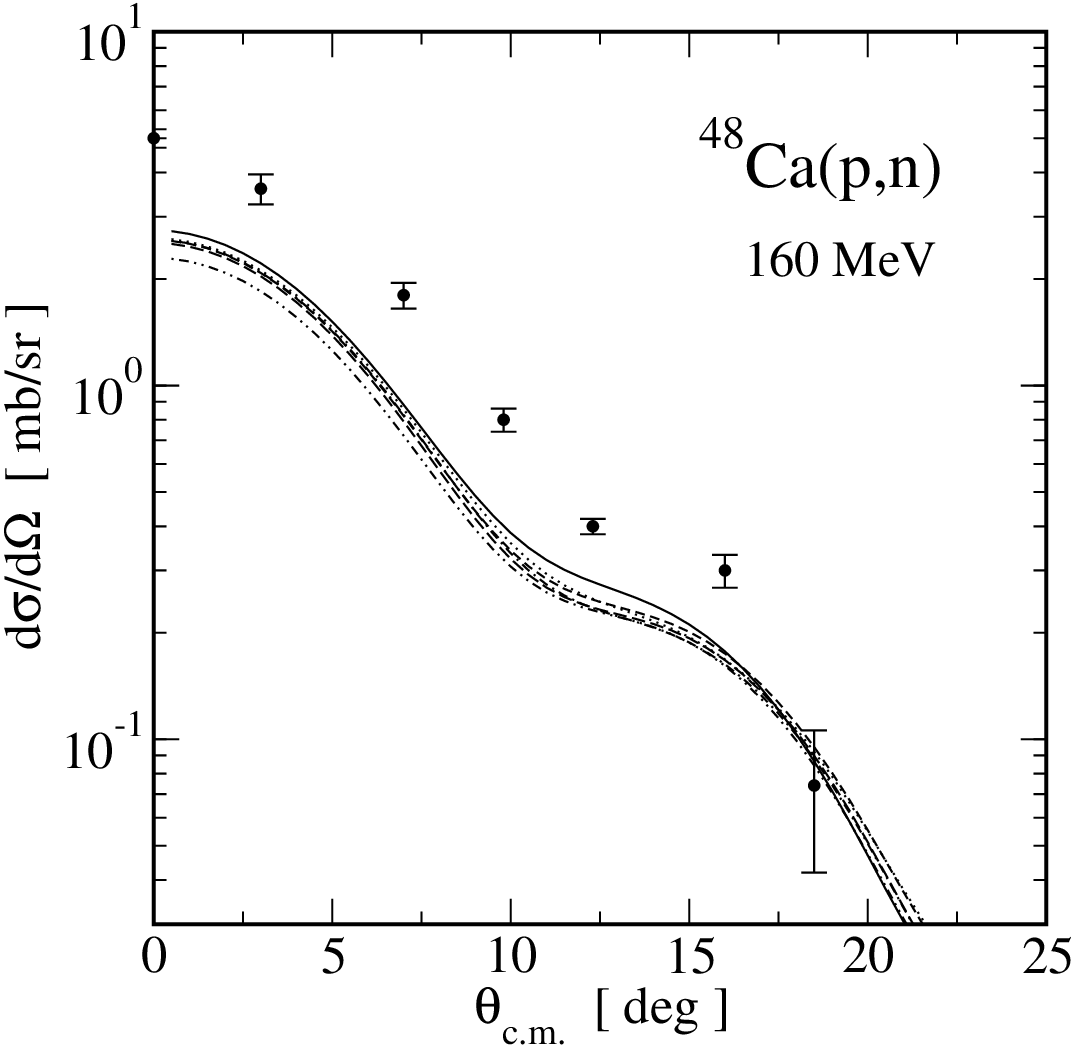} 
\medskip
\caption{\protect\small
\label{pnCa48_160}
         Measured \cite{And85} and calculated differential cross-section 
	 for $^{48}$Ca($p,n$) at 160 MeV.
	 The curve patterns follow the same convention as in
	 Fig. \ref{fermiDAT}. 
        }
\end{figure}

\begin{figure}[ht]
\includegraphics[scale=0.50,angle=00] {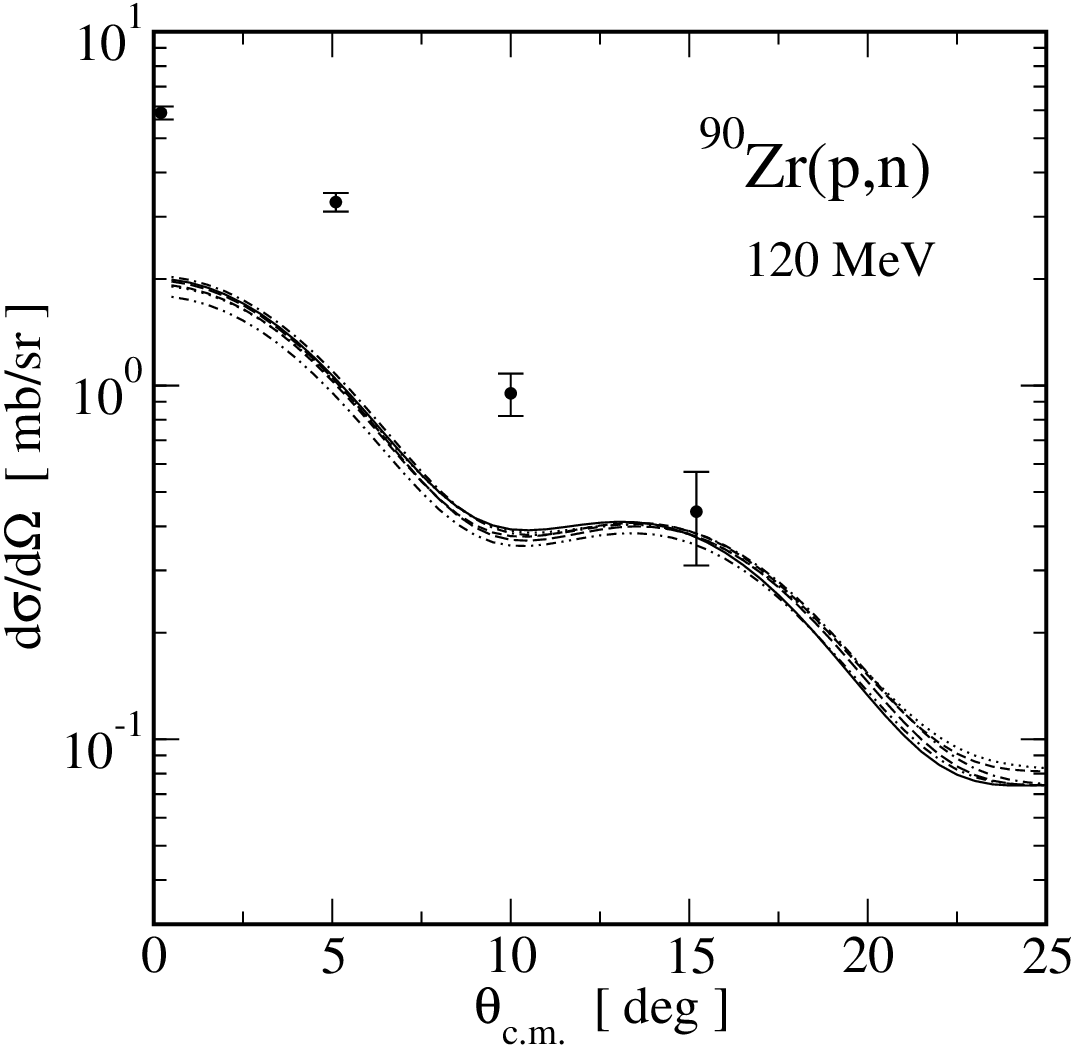} 
\medskip
\caption{\protect\small
\label{pnZr90_120}
         Measured \cite{Bai80} and calculated differential cross-section 
	 for $^{90}$Zr($p,n$) at 120 MeV.
	 The curve patterns follow the same convention as in
	 Fig. \ref{fermiDAT}. 
        }
\end{figure}

\begin{figure}[ht]
\includegraphics[scale=0.50,angle=00] {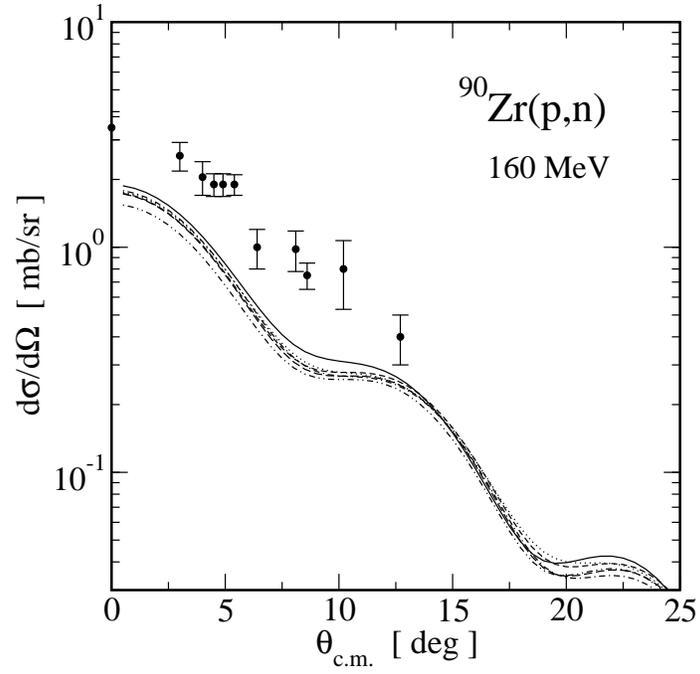} 
\medskip
\caption{\protect\small
\label{pnZr90_160}
         Measured \cite{Sug82} and calculated differential cross-section 
	 for $^{90}$Zr($p,n$) at 160 MeV.
	 The curve patterns follow the same convention as in
	 Fig. \ref{fermiDAT}. 
        }
\end{figure}

In addition to $^{14}$C(\emph{p,n}), we have also performed 
calculations for quasielastic nucleon charge-exchange scattering on
$^{48}$Ca ($E_x=$ 6.67 MeV \cite{And91}) and $^{90}$Zr 
($E_x=$ 5.1 MeV \cite{Bai80}) using the $g$-matrix full-folding approach.
In Figs. \ref{pnCa48_135} and \ref{pnCa48_160} we present results
for  $^{48}$Ca(\emph{p,n}) at 
135 and 160 MeV (data taken from Ref. \cite{And85}), 
respectively. 
Similarly, in Figs. \ref{pnZr90_120} and \ref{pnZr90_160} we show results
for $^{90}$Zr(\emph{p,n}) at 
120 (data taken from Ref. \cite{Bai80}) and 
160 (data taken from Ref. \cite{Sug82} )MeV.
The curve patterns follow the same convention as in Fig. \ref{pnC14_120}.

In contrast to the $^{14}$C(\emph{p,n}) case, the Fermi cross sections
for these four applications are significantly underestimated relative to 
the data. 
The calculated $\sigma_F$ for $^{48}$Ca at 160 MeV averages 
2.6($\pm$0.1) mb/sr, considering the six \emph{NN} potential models. 
At the same energy, but in the case of $^{90}$Zr, the average 
is 1.7($\pm$0.1) mb/sr. 
The reported measurements for $^{48}$Ca and $^{90}$Zr are  
5.0$\pm$0.3 mb/sr and 3.4$\pm$0.4 mb/sr, respectively, roughly twice as 
large as the calculated values. 
The measured-to-calculated ratio of $\sigma_F$ at the lower energies
increases to 2.3 ($^{48}$Ca at 135 MeV) and 3.0 ($^{90}$Zr at 120 MeV).

\begin{figure}[ht]
\includegraphics[scale=0.70,angle=-90] {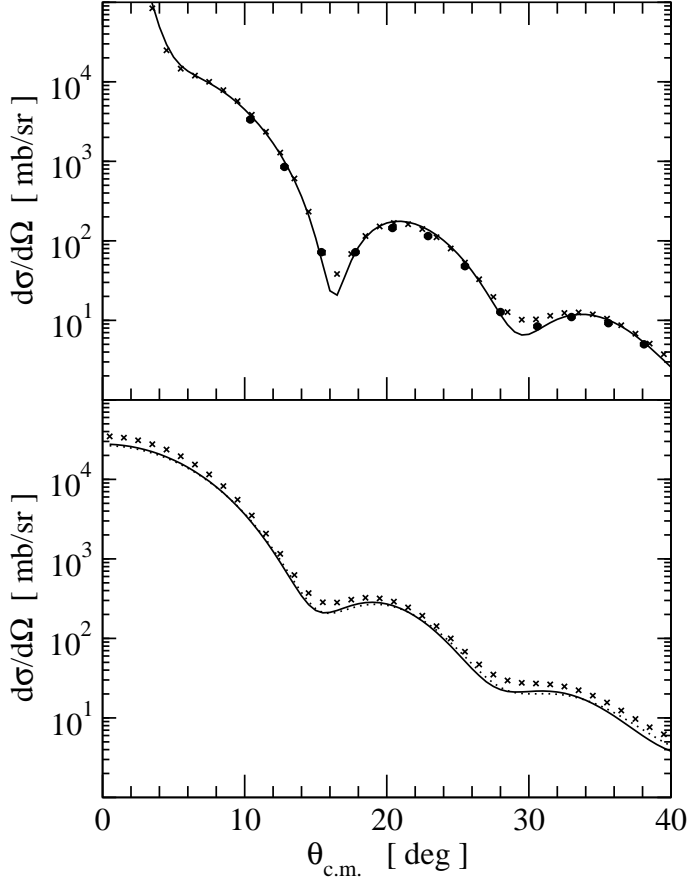} 
\medskip
\caption{\protect\small
\label{2Plots}
         Upper frame:
         Measured \cite{SCOTT} and calculated differential cross-section 
	 for proton elastic scattering from $^{90}$Zr at 121 MeV.
	 The solid curve corresponds to $g$-matrix full-folding OMP 
	 results, whereas the crosses correspond to a local WS 
	 parametrization.
	 Lower frame: The calculated \emph{uncharged} nucleon differential
	 cross-section based on $U_{pp}$ (solid curve), $U_{nn}$ 
	 (dotted curve) and the WS parametrization (sparse crosses).
        }
\end{figure}

The results shown in Figs. \ref{pnCa48_135}-\ref{pnZr90_160} correspond 
to fully 
consistent coupled-channel results involving explicitly the diagonal 
($U_{pp}$ and $U_{nn}$) and off-diagonal ($U_{x}$) optical potentials 
based on the full-folding approach. 
Thus, they illustrate the degree of consistency among the three 
components of the potential within the theory.
This is an important point to keep in mind, since it is possible 
to improve the description of the data with the use of the DWBA,
supplemented by suitable phenomenological potentials for the \emph{nn}
channel.
Indeed, in the upper frame of Fig. \ref{2Plots} we compare the 
differential cross section for proton scattering from $^{90}$Zr at 121 MeV.
The elastic scattering data are from Ref. \cite{SCOTT}, 
the solid curve represents the 
full-folding OMP results and the sparse crosses  
a standard Woods-Saxon (WS) parametrization.
Although the full-folding results are in superior agreement with the
data over the whole angular range, the quality of both descriptions
at angles below 25 degrees is comparable. 
A comparison at the hadronic level alone (i.e. suppressing the Coulomb
term) is presented in the
lower frame, where we show the differential cross 
section for \emph{uncharged} nucleon scattering using the full-folding 
$U_{pp}$ (solid curve), $U_{nn}$ (dashed curve) and the local WS 
parametrization (sparse crosses). 
The similarity between the $U_{pp}$ and $U_{nn}$ is striking,
suggesting their equivalence. 
The WS result is slightly larger than the other two cases.

\begin{figure}[ht]
\includegraphics[scale=0.70,angle=0] {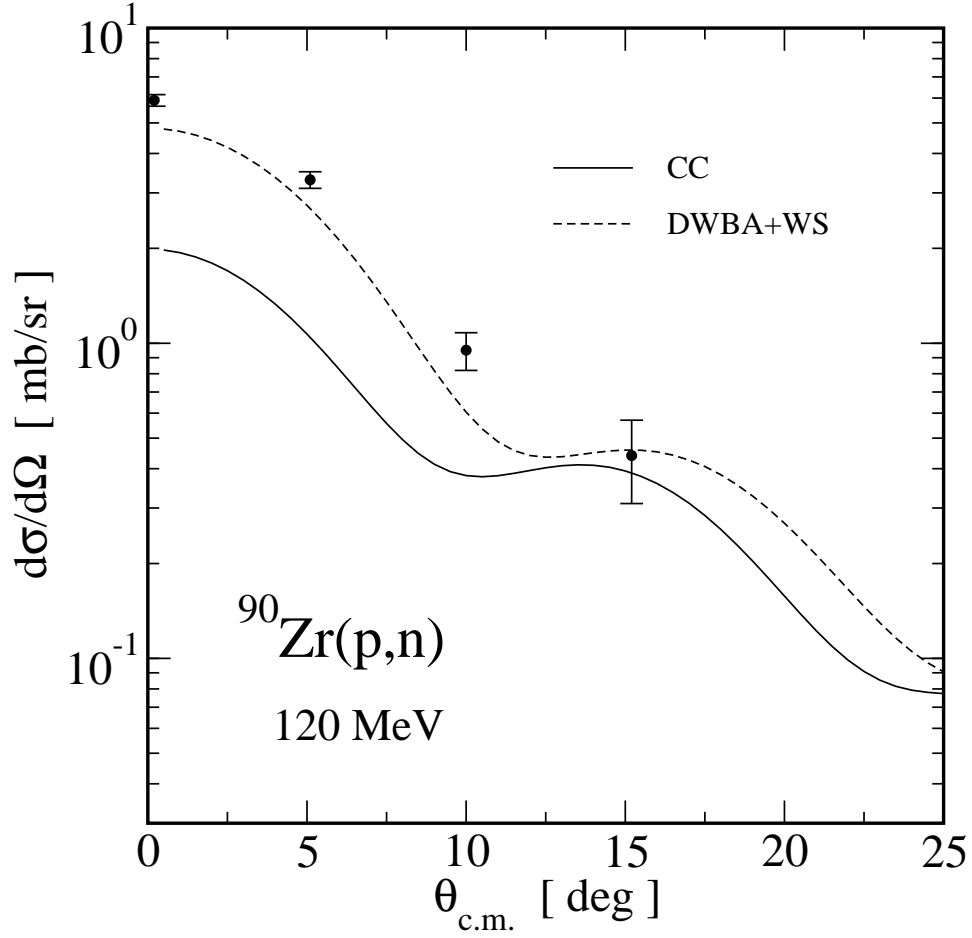} 
\medskip
\caption{\protect\small
\label{dwba}
         Measured \cite{Bai80} and calculated differential cross section 
         for $^{90}$Zr(\emph{p,n}) 
	 charge-exchange reaction based on full coupled-channel
	 (solid curve) and DWBA using the parametric WS potential
	 (dashed curve). 
        }
\end{figure}

In a DWBA application it is customary to use, for the scattered 
neutron wave, $U_{nn}:=U_{pp}$. 
In Fig. \ref{dwba} we compare the 
consistent coupled-channel full-folding OMP results (solid curve) 
with the corresponding DWBA results (dashed curves). 
In both cases we use the same $U_x$ based on the full-folding OMP.
However, for the DWBA application we use the parametric WS potential
already described.
The significant increase of the differential cross section
over whole the angular range is clear.

\begin{figure}[ht]
\includegraphics[scale=0.70,angle=-90] {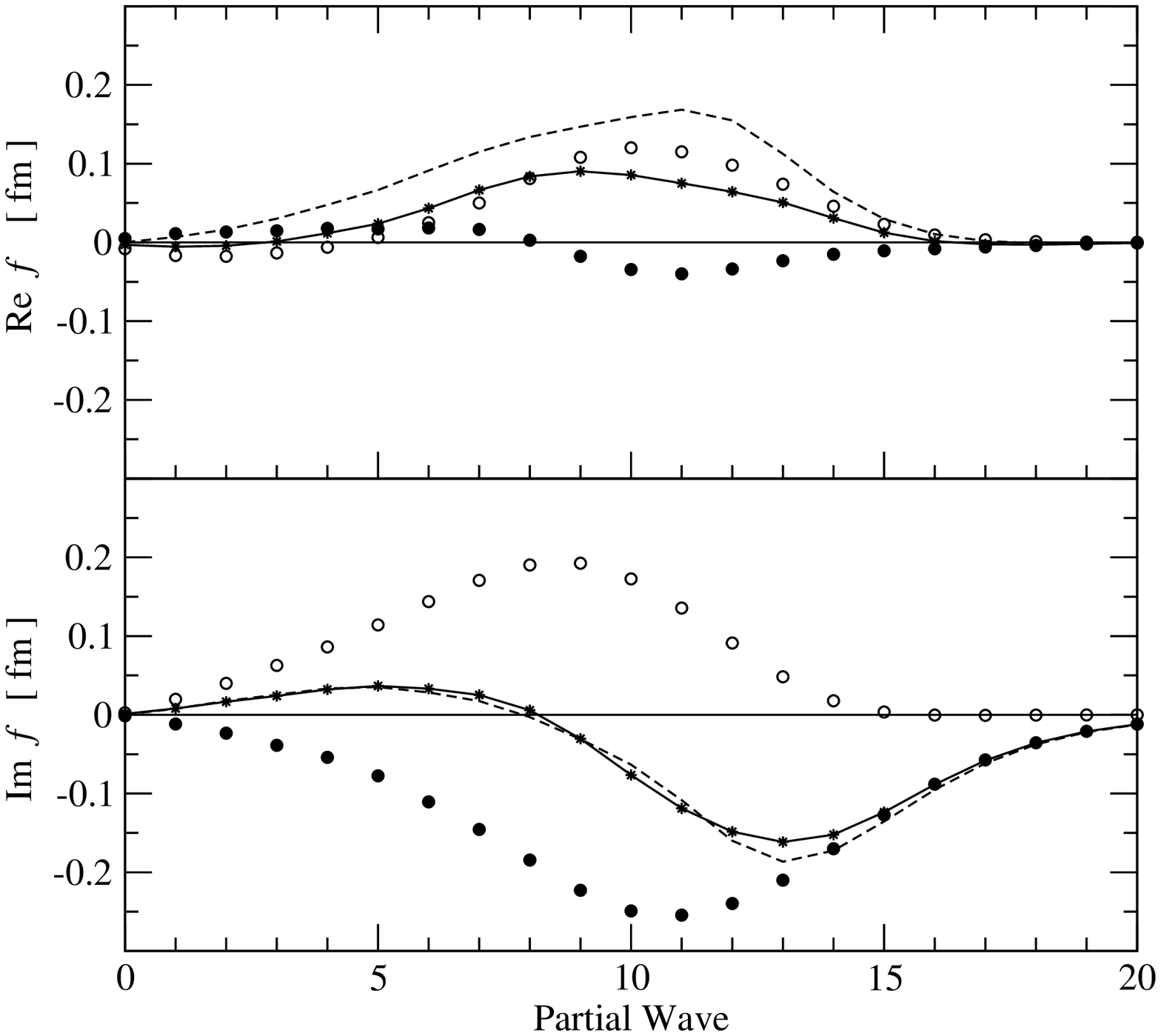} 
\medskip
\caption{\protect\small
\label{famplit}
         The real (upper frame) and imaginary (lower frame) component
         of the scattering amplitude $f$ as function of the orbital
         angular momentum. The solid curves, ${\bf \ast}$ symbols and dashed 
         curves represent the exact CC, DWBA-CC and DWBA-WS results, 
         respectively. The filled and empty circles represent the $U_{x}$ and
         $U_{nn}$ contributions to the total amplitude 
         [c.f. Eq. (\ref{amplit})]. 
        }
\end{figure}

In Fig. \ref{famplit} we present the scattering amplitude ($f$) as
function of the angular momentum $l$. The upper and lower frame
are for the real and imaginary components, respectively.
The filled (open) circles corresponds to exact coupled-channel results
extracted from Eqs. (\ref{amplit}) and (\ref{intamplit}), 
with the solid curves representing the sum of both contributions.
The stars ($\ast$) represent results based on DWBA using the
full-folding \emph{x} and \emph{nn} couplings (DWBA-CC).
The dashed curves correspond to DWBA calculations where the \emph{nn}
term has been replaced by a local WS parametrization (DWBA-WS).
Clearly the imaginary component of the amplitude (lower frame)
exhibits minor differences  among the exact coupled-channel, 
DWBA-CC and DWBA-WS results.
Additionally, we observe that the total amplitude is the result of a mutual
cancellation of the $U_x$ and $U_{nn}$ contributions.
In the upper frame for the real components, however, we notice major 
differences between DWBA-WS and either of the two coupled-channel 
applications.
Indeed, the real part of of the DWBA-WS amplitude is nearly twice
as large as the other two. 
This difference is exclusively due to the nature of the diagonal coupling, 
extracted phenomenologically from other sources. 
In that sense, a phenomenological improvement of the data is achieved
at the expense of a consistent two-channel description of the 
charge-exchange process.

The difficulties discussed above have been explored in the context of the
neutron-to-proton relative distribution.
We have investigated the discrepancy between the consistent coupled-channel
$g$-matrix OMP results and the measured quantities by studying the 
sensitivity of the scattering observables to variations in the neutron 
density \cite{Are05a}. 
Preliminary results along this line suggest the presence of a neutron 
halo in the nuclear periphery,
that is to say, pronounced ratio of the neutron to proton density 
away from the nuclear surface.
A more thorough investigation of this fact is a natural
extension of the present work, and should not be limited to the radial
matter distribution but include other features of the mixed density as well.
An additional element in this discussion is the charge-dependence of the
\emph{NN} interaction, which in this work, has only been included
in the $T_z=0$ \emph{NN} sector.

\subsection{Comparison with other results}

Various microscopic studies on (\emph{p,n}) charge-exchange reactions
have been presented in the past.
The most recent one by Bauge \emph{et al.} \cite{Bau01} consists of a 
Lane-consistent semi-microscopic optical model approach where an 
energy-dependent potential depth normalization factor is applied to the 
isovector components. 
These normalization factors are then used consistently in elastic and
quasielastic scattering.
Their quasielastic charge-exchange applications between 100 and 200 MeV
exhibit less pronounced discrepancies with the data relative to the ones 
presented in this work.
A microscopic extraction of these scaling factors remains to be seen.

A relativistic description of (\emph{p,n}) quasielastic reactions has
also been provided by Clark \emph{et al.} \cite{Cla84}. 
In their case the relativistic impulse approximation is used, where
Lorentz vector and scalar, isovector and isoscalar \emph{NA}
optical potentials are obtained from invariant \emph{NN} amplitudes.
Applications at 160 MeV for $^{90}$Zr\emph{(p,n)} lead to a
Fermi cross section weaker by a factor of 2 relative to the data.
Their interpretation is that the diagonal optical potential 
presents too much absorption.
A sizable improvement in the description of the differential
cross section is obtained with the use of phenomenological
(diagonal) potentials in conjunction with DWBA.

Although both approaches provide means by which to improve the
description of the data, they also emphasize the limited understanding
of this very simple process.

\section{Summary and conclusions}
We have studied quasielastic (\emph{p,n}) charge-exchange reactions
to the IAS at proton energies between 80 and 800 MeV.
The study is based on an extension of the microscopic full-folding
optical model approach for \emph{NA} elastic scattering
to processes where spin and isospin excitations become allowed.
Thus, we have developed the necessary spin-isospin recoupling 
in order to identify the off diagonal terms responsible for the 
exchange mechanism.
In this particular work we focus on isospin excitations only.
The resulting couplings, in the form of non-local optical potentials,
are then applied to quasielastic (\emph{p,n}) reactions
on $^{14}$C, $^{48}$Ca and $^{90}$Zr at intermediate proton energies.
Emphasis is given to the full consistency of the microscopic approach,
without resorting to the use of adjusting parameters of the model 
in order to describe the data. 
Indeed, it has been a primary focus of this work to disclose its 
limitations in its most complete current form, \emph{i.e.} with 
an accurate account of the Fermi motion (off shell), and implied 
non localities in the optical potentials.
Exact coupled-channel calculations, within numerical accuracy,
were performed to obtain the scattering observables.
Comparisons with the data show a persistent deficiency of the theory
to account for them in a consistent fashion.
This is particularly so in the case of the heavier targets,
where the zero-degree Fermi cross sections are substantially underestimated.
We observe, however, that the use of phenomenological representations of the
optical potential for the elastic channel yields sizable changes
of the Fermi cross sections. However, this phenomenological correction 
does not allow a clear identification of the missing microscopic effects 
needed to account for the data.

The above mentioned limitations become less dramatic in the case of
(\emph{p,n}) charge-exchange reactions on $^{14}$C. In this case
the full-folding optical model approach is able to explain reasonably 
well the differential cross section at 120 MeV. 
Additionally, depending on the \emph{NN} potential model, the Fermi 
cross section is described with varying degrees of success in the
energy range 80-800 MeV. These applications represent extrapolations 
of the of the underlying \emph{NN} potential model and favor the
Nijmegen II and Argonne AV18 models. 
A more definitive assessment of such results would require, however, 
the account for the above-threshold inelasticities at the \emph{NN} 
level, an issue to be considered in the future.

Though this work represents the most complete microscopic calculations 
to date, accounting for full Fermi motion effects and implied non
localities, with an accurate treatment of the resulting coupled-channel
reaction equations, we conclude that the model is still unable to 
satisfactorily describe the data.
The discrepancies are not minor, especially in the cases of
(\emph{p,n}) reactions on $^{48}$Ca and $^{90}$Zr studied here, 
where the zero-degree Fermi cross section is underestimated by 
factors of 2 or 3.
Although the data can be described by using distorted waves based on 
phenomenological optical models, or by adjusting other parameters of the model, 
we have preferred to maintain the formal structure of the theory to  
stress the limitations of the microscopic model in its current form.
There are other venues for the inclusion of additional effects not 
discussed here. Among the most immediate ones we mention the role 
of asymmetric nuclear matter in the \emph{NN} effective interaction 
and the neutron distribution in the target ground state \cite{Are05a}.
Work toward inclusion of isospin asymmetry in the effective 
interaction is under way.

\begin{acknowledgments}
The authors thank Prof. Jacob Rapaport for his kindness in providing
available charge-exchange data.
H. F. A. thanks the hospitality of the Department of Physics and Astronomy 
of the University of Georgia during his visit in the course of this
collaboration.
Partial funding for this work has been provided by FONDECYT under
Grant No 1040938.
\end{acknowledgments}

\newpage



\newpage

\newpage

\newpage

\newpage

\newpage

\newpage

\newpage

\newpage

\newpage

\newpage


\begin{thebibliography}{aaa}
%
\bibitem{Bao05}   
    Bao-An Li, Pawel Danielewicz, and William G. Lynch,
    Phys. Rev. C 71, 054603 (2005).
\bibitem{Dal05}
    E. N. E. van Dalen, C. Fuchs, and Amand Faessler,
    Phys. Rev. C 72, 065803 (2005).
\bibitem{Ron06}
    Jian Rong, Zhong-Yu Ma, and Nguyen Van Giai,
    Phys. Rev. C 73, 014614 (2006).
\bibitem{Rik02}
    J. R. Stone, P. D. Stevenson, J. C. Miller, and M. R. Strayer,
    Phys. Rev. C 65, 064312 (2002).
\bibitem{Sch00}
    H. Scheit, F. Mar\'echal, T. Glasmacher, E. Bauge, Y. Blumenfeld, 
    J. P. Delaroche, M. Girod, R. W. Ibbotson, K. W. Kemper, J. Libert, 
    B. Pritychenko, and T. Suomij\"arvi,
    Phys. Rev. C 63, 014604 (2001).
\bibitem{Noj96}
    R. Nojarov,
    Phys. Rev. C 54, 668 (1996).
\bibitem{Are02}
    H. F. Arellano and H. V. von Geramb,
    Phys. Rev.C 66, 024602-1 (2002).
\bibitem{Are89}
    H. F. Arellano, F. A. Brieva, and W. G. Love,
    Phys. Rev. Lett. 63, 605 (1989).
\bibitem{Are95}
    H. F. Arellano, F. A. Brieva, and W. G. Love,
    Phys. Rev. C 52, 301 (1995).
\bibitem{Amo00} 
    K.~Amos, P.~J. Dortmans, H.~V. von Geramb, S.~Karataglidis, and J.~Raynal, 
    Adv. in Nucl. Phys. 25 (2000) 275.
\bibitem{Che92}
    Taksu Cheon and Kazuo Takayanagi,
    Phys. Rev. Lett. 68, 1291 (1992).
\bibitem{Cla84}
    B. C. Clark, S. Hama, E. Sugarbaker, M. A. Franey, R. L. Mercer, 
    L. Ray, G. W. Hoffmann, and B. D. Serot,
    Phys. Rev. C 30, 314 (1984).
\bibitem{Are90b}
    H.F. Arellano, F.A. Brieva and W.G. Love,
    Phys. Rev. C 42, 652 (1990).
\bibitem{Lan62}
    A. M. Lane, 
    Nucl. Phys. 35, 676 (1962).
\bibitem{Hof73}
    G. W. Hoffmann,
    Phys. Rev. C 8, 761 (1973).
\bibitem{Lac80}
    M. Lacombe, B. Loiseau, J. M. Richard, and R. Vinh Mau,
    J. C\^ot\'e, P. Pir\'es and R. de Tourreil,
    Phys. Rev. C 21, 861 (1980).
\bibitem{Sto94}
    V. G. J. Stoks, R. A. M. Klomp, C. P. F. Terheggen, and J. J. de Swart,
    Phys. Rev. C 49, 2950 (1994).
\bibitem{Wir95}
    R. B. Wiringa, V. G. J. Stoks, and R. Schiavilla,
    Phys.~Rev. C 51, 38 (1995).
\bibitem{Mac01}
    R. Machleidt,
    Phys. Rev. C 63, 024001 (2001).
\bibitem{Are94}
    H. F. Arellano, F. A. Brieva, and W. G. Love,
    Phys. Rev. C 50, 2480 (1994).
\bibitem{Jag74}
    C. W. de Jager, H. de Vries, and C. de Vries,
    At. Data Nucl. Data Tables 14, 479 (1974).
\bibitem{Rap83}
    Jacob Rapaport, unpublished (private communication).
\bibitem{And91}
    B. D. Anderson, M. Mostajabodda'vati, C. Lebo, 
    R. J. McCarthy, L. Garcia, J. W. Watson, and R. Madey,
    Phys. Rev. C 43, 1630 (1991). 
\bibitem{Tad82}
    T. N. Taddeucci, J. Rapaport, D. E. Bainum, C. D. Goodman, C. C. Foster, C. Gaarde, J. Larsen, C. A. Goulding, D. J. Horen, T. Masterson, and E. Sugarbaker,
    Phys. Rev. C 25, 1094 (1982), and references therein.
\bibitem{Sug90}
    E. Sugarbaker, D. Marchlenski, T. N. Taddeucci, L. J. Rybarcyk, 
    J. B. McClelland, T. A. Carey, R. C. Byrd, C. D. Goodman, W. Huang, 
    J. Rapaport, D. Mercer, D. Prout, W. P. Alford, E. 
G\"ulmez, 
    C. A. Whitten, and D. Ciskowski,
    Phys. Rev. Lett. 65, 551 (1990).
\bibitem{Bai80}
    D. E. Bainum, J. Rapaport, C. D. Goodman, D. J. Horen, C. C. Foster, 
    M. B. Greenfield, and C. A. Goulding,
    Phys. Rev. Lett. 44, 1751 (1980).
\bibitem{And85}
    B. D. Anderson, T. Chittrakarn, A. R. Baldwin, C. Lebo, R. Madey, 
    P. C. Tandy, J. W. Watson, B. A. Brown, and C. C. Foster,
    Phys. Rev. C 31, 1161 (1985).
\bibitem{Sug82}
    E. Sugarbaker {\em et al.},
    Proceedings of the International Conference on Nuclear Structure,
    Amsterdam, 1982, 
    edited by A. Van Der Wonde and B. J. Verhaar, pp. 77.
    Jacob Rapaport (private communication).
\bibitem{SCOTT}
    Alan Scott (unpublished).
\bibitem{Are05a}
    H. F. Arellano and W. G. Love,
    Nucl. Phys. A 755C, 527 (2005).
\bibitem{Bau01}
    E. Bauge, J. P. Delaroche, and M. Girod,
    Phys. Rev. C 63, 024607 (2001)
\end{thebibliography}
\end{document}